\documentstyle[preprint,aps,epsf]{revtex}

\begin{document}

\author{Nara Guisoni\thanks{
Electronic address: nara@fge.if.usp.br} \\
Vera Bohomoletz Henriques\thanks{Eletronic address:
vhenriques@if.usp.br} \\ }

\title{A geometric model for cold water and liquid-liquid transitions}

\address{Instituto de F\'{\i}sica, Universidade de S\~ao Paulo, \\
C.P. 66318, cep 05315-970, S\~ao Paulo, SP, Brazil}

\maketitle

\begin{abstract}
Water is an associated liquid
in which the main intermolecular interaction is the hydrogen bond (HB) which is
limited to four per atom, independently of the number of neighbours. We
have considered a hydrogen bond net superposed on  Bernal's geometric model
for liquids, which allows for different local environments for the
liquid particles. In this study, a mean-field treatment of the
two-dimensinal version of the model is discussed. 
Under pressure the model exhibits three phases of different densities and a
coexistence line ending in a critical point between low and high density
phases. Entropy of the HB network plays an essential role in defining the
slope of the coexistence line. The model
behaviour might be of interest in describing supercooled water and
liquid-liquid transitions of other substances.

\end{abstract}

\newpage

\section{Introduction}
Liquid polymorphism has been proposed for different systems, both from experimental 
evidence, as well as from theoretical arguments
\cite{poole-grande,mishima-stanley,exp-c,ferraz,glosli,thiel,phosphorus,aasland,tanaka,tejero}. The relevant systems must
have open molecular coordination environments at low pressures, and may
either be liquids associated through a hydrogen-bond net, water being the
main example \cite{mishima-stanley}, or substances for which the different
possibilities for hybridyzed eletronic orbitals imply structures of
different densities, as in the case of carbon
\cite{exp-c,ferraz,glosli,thiel}. 

The anomalous behaviour of water has
intrigued physicists and chemists for a long
time \cite{franks,kauzmann,russo,stillinger}. It is presently accepted that
the special properties are related to hydrogen bonds. Because the energy involved in a
hydrogen bond is an order of magnitude larger than the typical van der Waals 
energy \cite{kauzmann,egelstaff}, the fusion and boiling temperatures of
water are much higher than those of homologous substances \cite{stillinger}. In fact, liquid
water exhibits an extense random network of hydrogen bonds which  
continuously reformulates, but presents local tetrahedral
symmetry, responsible for the large volume at low
temperatures \cite{stillinger,teixeira}.

In the last years, interest in supercooled water arose
\cite{mishima-stanley,stanley2,belli,belli2} as a means to explore the  
behaviour of isothermal compressibility and specific heat which, differently
from most substances, rise as one lowers temperature towards the fusion
point and beyond \cite{kauzmann,spinodal}. Search for an explanation of
these two properties led to three different hypotheses: a retracing liquid-gas
spinodal \cite{spinodal}, a second critical point in the
supercooled region \cite{mishima-stanley,stanley2,tc2}, and a
``singularity-free" scenario \cite{sastry,sastry-model,percolacao}, in
opposition to the first two. Sastry and collaborators \cite{sastry}
have shown that a singularity is not the only explanation for the
compressibility or specific heat low temperature behaviour, but molecular
dynamics simulations of ST2 water
\cite{sc} indicate that
a divergence of compressibility could occur around 2000atm, 230K . 
Among the several simplified models presented in the literature, some do
\cite{debe,pony,poole}, other do not
\cite{sastry,sastry-model,percolacao} present the second critical point, in
some cases depending on model parameters used. From
the experimental point of view, the question  is very difficult to settle,
because in the supercooled region one encounters a nucleation line 
($-38^oC$, at 1 atm) beyond which the system spontaneously crystallizes
\cite{mishima-stanley}.
  
The second critical point of water would be the end of a coexistence line
between two metastable liquids with different densities (HDL and LDL for
high and low density liquid) \cite{mishima-stanley}. These two liquids would be related to two
amorphous solids which exhibit coexistence and are found experimentally beyond the nucleation
line \cite{mishima}. The existence of the amorphous phases lends
support to the idea of a liquid-liquid transition.

The idea of liquid polymorphism has been applied not only to water,
but also to certain molecular liquids, such as I, Se and S, in order to explain
an abrupt increase in conductivity, which would be related to a first
order phase transition  in the supercooled region
\cite{poole-grande}. 
Experimental evidence of a transition 
between two distinct thermodynamically {\it stable} liquid phases
in carbon was found \cite{exp-c} some years after a
suggestion of its existence appeared in the literature\cite{ferraz}. This
transition has also been reported from 
molecular dynamics simulations \cite{glosli} and a scaling
formalism \cite{thiel}. Experimental results also report a liquid-liquid
transition in phosphorus in the stable phase \cite{phosphorus}.
 
A few decades ago Bernal proposed a model for liquids which consisted of
a close-packed assembly of different regular or quasi-regular polyhedra
\cite{bernal}. The typical liquid configuration was expected to be
statistically homogeneous and to possess no long-range order, due to the
presence of different kinds of polyhedra. Molecules would be located only at
the vertices and the edges would represent an average nearest neighbour
intermolecular distance. The absence of holes, one of the model hypotheses,
reduces the number of allowed polyhedra. Collins studied a mean-field
solution for a version of Bernal's model in two dimensions \cite{collins}, in
which case the polygons to be considered are squares and triangles. Collins
demonstrated that the model could exhibit a phase transition, at fixed
volume, for a specially designed set of interaction constants with no
physical motivation. The author was looking for a melting transition, but
recognized that the phase change was more analogous to a change of
association number from one liquid to another. Interestingly, a recent
study \cite{tiling} of a dense system of repulsive Lennard-Jones
particles showed that the spatial configurations can be interpreted in terms
of random square-triangle tilings, as in Bernal's proposal. A transition
between two densities was also found.
 
We have considered the spatial distribution of
particles proposed by Collins and introduced interactions between the
molecules which would mimic the HB network. Thus the distribution of energies
is a result of the disponibilities of molecules to engage in HBs and
favours low-density arrangements at low temperatures and pressures.  Also, besides the
translational entropy for simple liquids, an entropy term related to the
distribution of HBs must be considered. The model does not allow for
vacancies, so a gaseous phase is disregarded. A mean-field treatment of the
system behaviour as a function of pressure and temperature is undertaken. We
have also considered a  variant of the model, which could describe
non-associated liquids with preference for low density at low pressures (that
would be the case of carbon). In the latter case, the HB entropy
contribution is absent.

Polymorphism and a critical point are found for both models. However, the
slope of the coexistence lines depends on competing entropy contributions. The
model for water and the corresponding phase diagram is presented in section
II. Liquid-liquid transitions and the phase diagram for non-associated
liquids are discussed  in section III. Discussion and summary are presented in section IV.

\section{The model for water}

\subsection{The geometric description}

\begin{figure}[th]

\epsfysize=8.77cm \epsfxsize=12.64cm

\epsffile{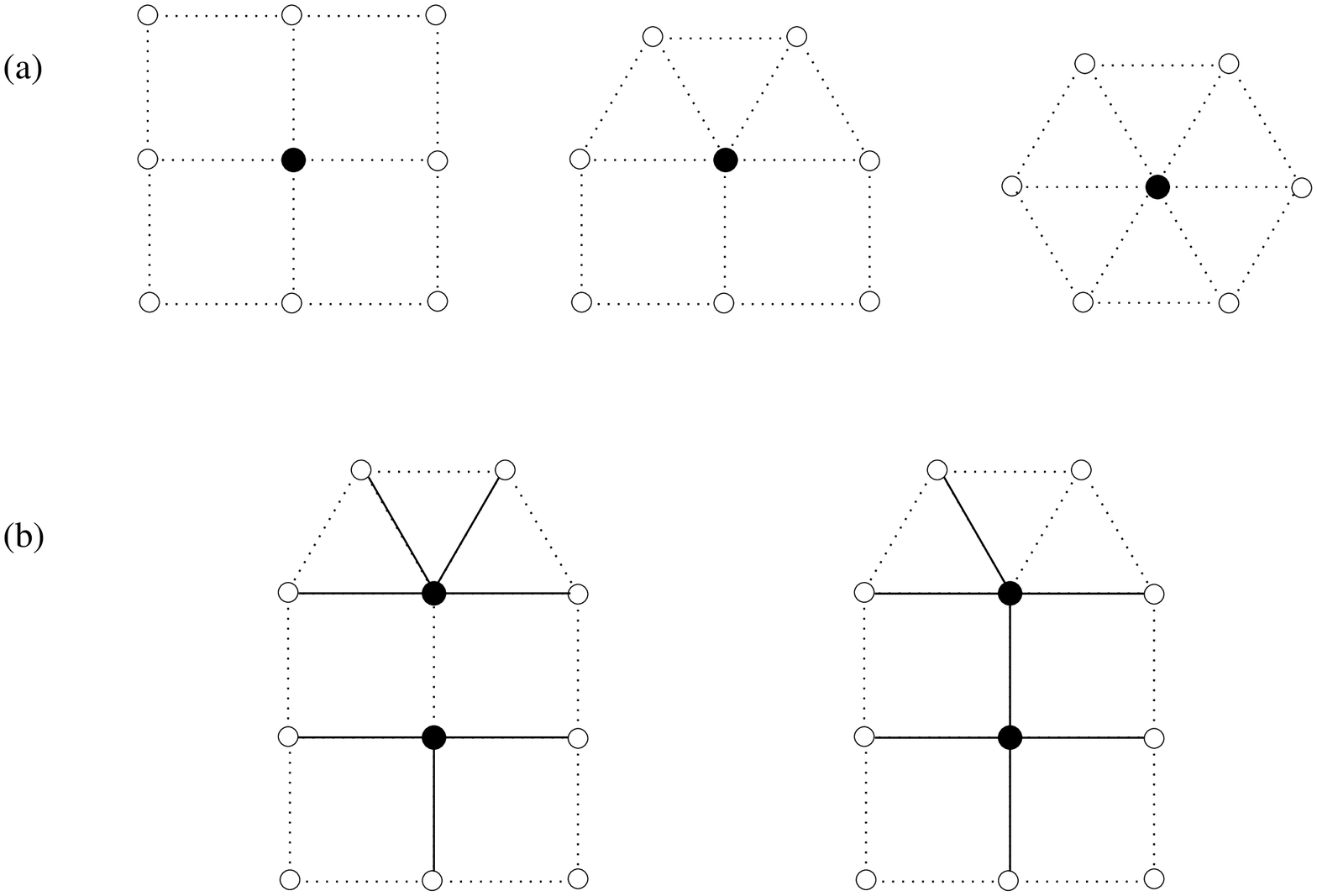}

\vspace*{3cm}

\caption{\footnotesize (a) Three possibilities for the local molecular
environment: $4$, $5$ and $6$-molecules (black dots).
(b) The existence of an HB between a $4$ and a $5$ molecule
depends on the distribution of the four molecular bonds (full line)
over the respective
``neighbouring" lines (dashed lines).
In the left figure no HB occurs between the two central molecules
because the $5$-molecule has engaged its four bonds with the other 4
neighbours.\label{desenho}}


\end{figure}

In two dimensions, the Bernal liquid consists of a system of adjacent and
randomly distributed squares and triangles of equal sides.  The particles
are localized at the vertices and there are three possibilities for
the number
of nearest neighbours $r$, as showed in Fig. \ref{desenho}-a.
A $4$ and a $6$-molecule can never be nearest neighbours by a
geometric constraint.
We call $n_r$ the number fraction of molecules with coordination $r$ ($r$-molecule) 
and v$_r$ the corresponding specific volume, 
\[
v_4=b^2 
\]

\begin{equation}
v_5=b^2(2+\sqrt{3})/4
\end{equation}

\[
v_6=b^2\sqrt{3}/2 
\]
for fixed intermolecular lenth $b$. 

Writing $v=V/N$ for the volume per molecule, number and volume 
conservation are written as:
\begin{equation}
\sum_{r=4,5,6}n_r=1  \label{numero}
\end{equation}

\begin{equation}
\sum_{r=4,5,6}n_rv_r=v.  \label{volume}
\end{equation}

Interactions must still be specified. In simple liquids, van der Waals
interactions depend mainly on interatomic spacings. 
Hydrogen bonding depends on disponibility of neighbours to accept or
donate hydrogens, thus in liquid water loss of translational
order and increased density results in the frustration of some of the HBs 
for molecules with over four nearest neighbours \cite{stillinger}. In  order to
represent this property we introduce, alongside with the distribution of
triangles and squares, a distribution of bonds. We have considered four as
the maximum number of bonds per molecule \cite{foot}, thus the HB between a
$4-5$ pair, for instance, may be absent, depending on the distribution of bonds
among the neighbours of the $5$-molecule (see
Fig. \ref{desenho}-b). Note that the energy of
the pair depends on the distribution of HBs amongst their neighbours. 
Because the hydrogen bond is an order of magnitude larger than van der Waals interaction
\cite{kauzmann}, in this study we have considered
only the HB interaction. The directionality of the hydrogen bond was
also ignored.

\subsection{A mean-field approach for the HB network}

Our approximation for the distribution of HBs
is the following: the four possible bonds of an $r$-molecule are distributed
randomly over the $r$ possible lines and the average energy of an $r-s$ pair is
calculated as resulting from the two independent distributions (for
example: for $5$-molecules there are five possible distributions of four bonds
on five lines, so the probability of an HB for a $5-5$ pair is $\frac{16}{25}$).
In this manner, if the HB energy is $-\epsilon$, then the average $r-s$
energy per bond is given by: $\Phi _{rs}=-\frac{4^2}{rs} \epsilon$ which makes
$\Phi_{44}$ (the energy of a pair of $4$-coordinated particles) the
minimum energy. Under this assumption, the energy of the model may be
written as:  $E=\sum_{rs}^{}N_{rs}\Phi _{rs}$
where $N_{rs}$ is the number of $rs$ pairs.

Standard mean-field conditions may then be implemented by assuming 
$N_{rs}=r N_r p_s(r) + s N_s p_r(s)$, where $N_r=n_rN$ is the total
number of $r$-molecules and $p_s(r)$ is the probability that an
$r$-molecule has an $s$-molecule as neighbour. Notice that the geometric
restriction must be taken into account, so assuming random
distribution of molecules we have:
\begin{eqnarray}
p_s(r)=\frac{N_s}{N_{tot}(r)}, 
\end{eqnarray}
where $N_{tot}(r)$ is the maximum number of neigbours for an
$r$-molecule ($N_{tot}(4)=N_4 + N_5$, $N_{tot}(5)=N_4 + N_5 + N_6$ and $N_{tot}(6)=N_5 + N_6$).

The average energy per molecule is then:
\begin{eqnarray}
e\left( \{n_i\} \right)=\frac12 \sum _r r n_r \sum _s p_s(r) \Phi _{rs},
\label{ener}
\end{eqnarray}
where $\frac12$ stands for double counting.

\subsection{Entropy and Gibbs free-energy} 
 
The partion function of the system in the Gibbs (constant pressure) ensemble is
written as:
\begin{eqnarray}
Z(T,P)=\sum _{\{n_i\}}'
\Omega(\{n_i\}) e^{-\frac{[E+PV]}{kT}} ,
\label{z}
\end{eqnarray}
where $\sum '$ represents the summation over $n_i$ constrained by
particle number conservation (Eq. \ref{numero}), and  $\Omega$
is the number of states with volume $V(\{n_i\})=Nv(\{n_i\})$ and energy
$E(\{n_i\})=Ne(\{n_i\})$ (Eqs. \ref{volume} and \ref{ener}). 
Two factors contribute to entropy, one related with the spatial
distribuition of particles, $\Omega_{part}(\{n_i\})=\frac{N!}{N_4!N_5!N_6!}$, and
the other related with the distribuition of HBs, $\Omega_{HB}(\{n_i\})$. 
The degeneracy of HBs is taken into account in the simplest way, by
considering independent molecules: 
\begin{eqnarray}
\Omega_{HB}=\prod _r D_r^{n_r} \label{s-ph},  
\end{eqnarray}
where $D_r=\frac{r!}{4!(r-4)!}$ is the number of possible arrangements
of four HBs over the vertices of  an $r$-molecule. Thus Z (Eq. \ref{z}) is
rewritten as:
\begin{eqnarray}
Z(T,P)=\sum_{\{n_i\}}'  e^{-
\frac{N\gamma (\{n_i\})}{kT}}, 
\end{eqnarray}
where $N\gamma =E-TS+PV$ and $S=k\ln \Omega _{tot}$ ($\Omega _{tot}=\Omega
_{part} \Omega _{HB}$). 
In the thermodynamic limit the Gibbs free energy per molecule is determined
by the equation:  $g(T,P)=\gamma (T,P;\{n_i\}_{\min })$, where
$\{n_i\}_{\min }$ is the set that minimizes $\gamma $. In order to find it
we minimized $\gamma $ with the constraint of
Eq. \ref{numero}, obtaining two transcendental equations, which are
solved numerically.
Henceforth the temperature and
pressure will be written in reduced units $t$ and $p$ ($K_BT/\epsilon$ and $Pb^2/\epsilon$, respectivelly).

\subsection{The phase-diagram}

We find the equilibrium values of $n_i$ as a function of
temperature and pressure using a combination of numerical
methods. A rough determination of the free energy
minima in an extensive search was followed by the Newton method, for more
precision.

\begin{figure}[th]

\epsfysize=12.77cm \epsfxsize=12.64cm

\epsffile{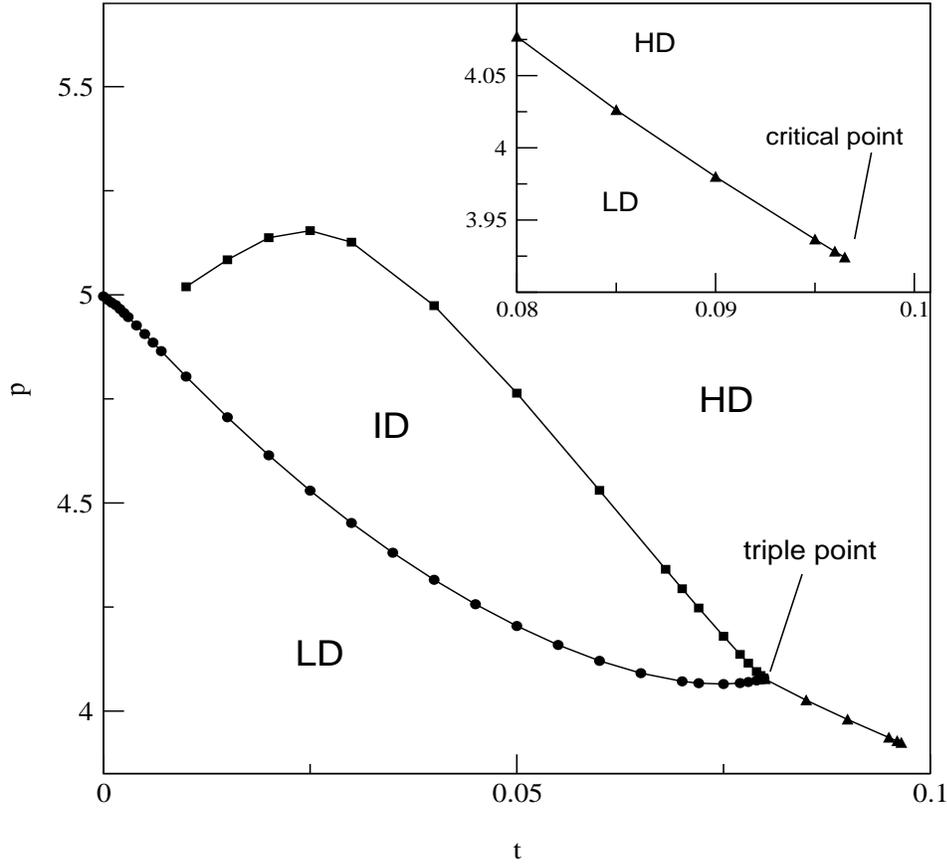}

\vspace*{3cm}

\caption{\footnotesize Coexistence lines and pressure-temperature phase
diagram in reduced variables. Three phases of low, intermediate and high density 
are identified (LD, ID and HD).
The LD phase has $n_4 > n_6 > n_5$ for low temperatures,
$n_6 > n_4 > n_5$ for $t>0.06$
and $n_6 > n_5 > n_4$ above $t=0.09$. ID and HD have, respectively $n_6 > n_4
> n_5$  and $n_6 > n_5 > n_4$, independently of temperature. The triple and
the critical points are indicated. The inset shows the region of interest
for supercooled water. \label{coex}}


\end{figure}

The phase diagram of the model for liquid water on the pressure-temperature
plane is shown in Fig. \ref{coex}. At lower temperatures, the model presents
two coexistence lines between three phases with different densities, with a
triple point at an intermediate temperature. At higher temperatures,
the coexistence between a high density (HD) and a low density (LD) phase ends
at a critical point. The critical temperature is obtained both from the
study of the free-energy surface as well from the evolution of the
order parameter defined as
$\lambda =\rho _{HD}-\rho _{LD}$ 
where $\rho _{HD,LD}$ are the corresponding densities.

The triple and
critical points (in reduced units) are given respectivelly by: $t_{tp}=0.0799$, $p_{tp}=4.0780$,
$t_{c}=0.0965$ and $p_{c}=3.9241$. Curiously, if one assumes $20
kJ/mol$ \cite{franks} for the HB energy ($\epsilon$), a critical
temperature of $232 K$ is obtained (against $230 K$ obtained from MD simulations \cite{sc}). As for the critical pressure, dimensionality would
enter the calculation, rendering it meaningless.

\begin{figure}[th]

\epsfysize=12.77cm \epsfxsize=12.64cm

\epsffile{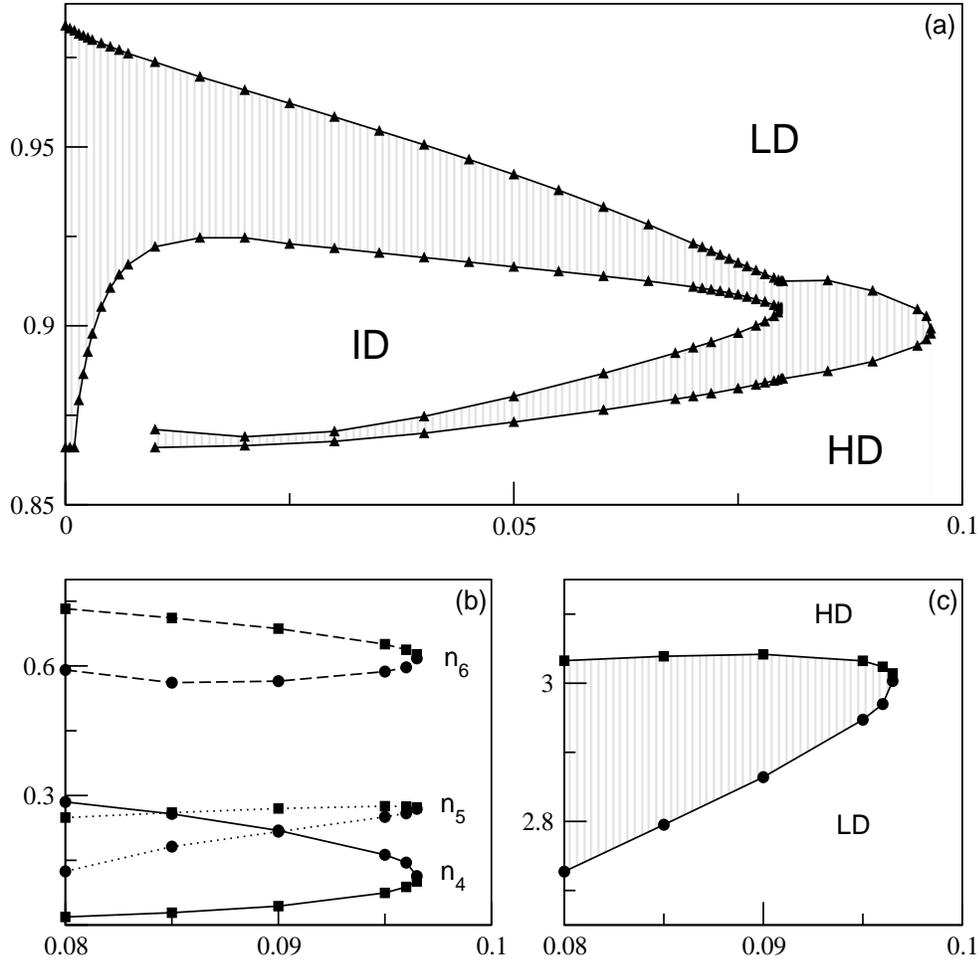}

\vspace*{3cm}

\caption{\footnotesize  (a) Volume versus temperature on the
coexistence lines. Coexistence regions are indicated by gray lines.
LD, ID and HD as in Fig. 2. 
(b) $n_4$ (full lines), $n_5$ (dashed lines) and $n_6$ (long
dashed lines) versus temperature for the LD (circles) and HD (squares) phases
on the coexistence line above the triple point. For the region of temperatures shown six-coordinated
particles predominate in both phases. 
(c) Entropy of the HD (squares) and LD (circles) phases versus
temperature on the coexistence line. \label{multi}}


\end{figure}

Fig. \ref{multi}-a illustrates discontinuity of volume on the coexistence
lines. The low density phase has more
four-coordinated molecules at low temperatures, but on the coexistence line
between LD and HD six-coordinated particles predominate in both phases, as can
be seen in Fig. \ref{multi}-b. The three phases differ basically in density.

The negative sign of the slope between the HD and LD phases in Fig.
\ref{coex} indicates, in
accordance with Le Chatelier's principle, that entropy increases while 
volume decreases, on transition into the high temperature
phase (HD), as can be seen in Figs.
\ref{multi}-a and \ref{multi}-c. The higher density phase has higher entropy
than the LD phase. Although it is true that the completely ordered $6$-phase has
higher entropy than the completely ordered $4$-phase, because of entropy of
bonds (see Eq. \ref{s-ph}), at finite temperatures competition with
translational entropy may invert the balance in some regions of the phase
diagram, as can be seen in Fig. \ref{coex} at very small temperatures. But the
overall picture is that the higher density phases also have higher entropy.

\begin{figure}[th]

\epsfysize=13.77cm \epsfxsize=12.64cm

\epsffile{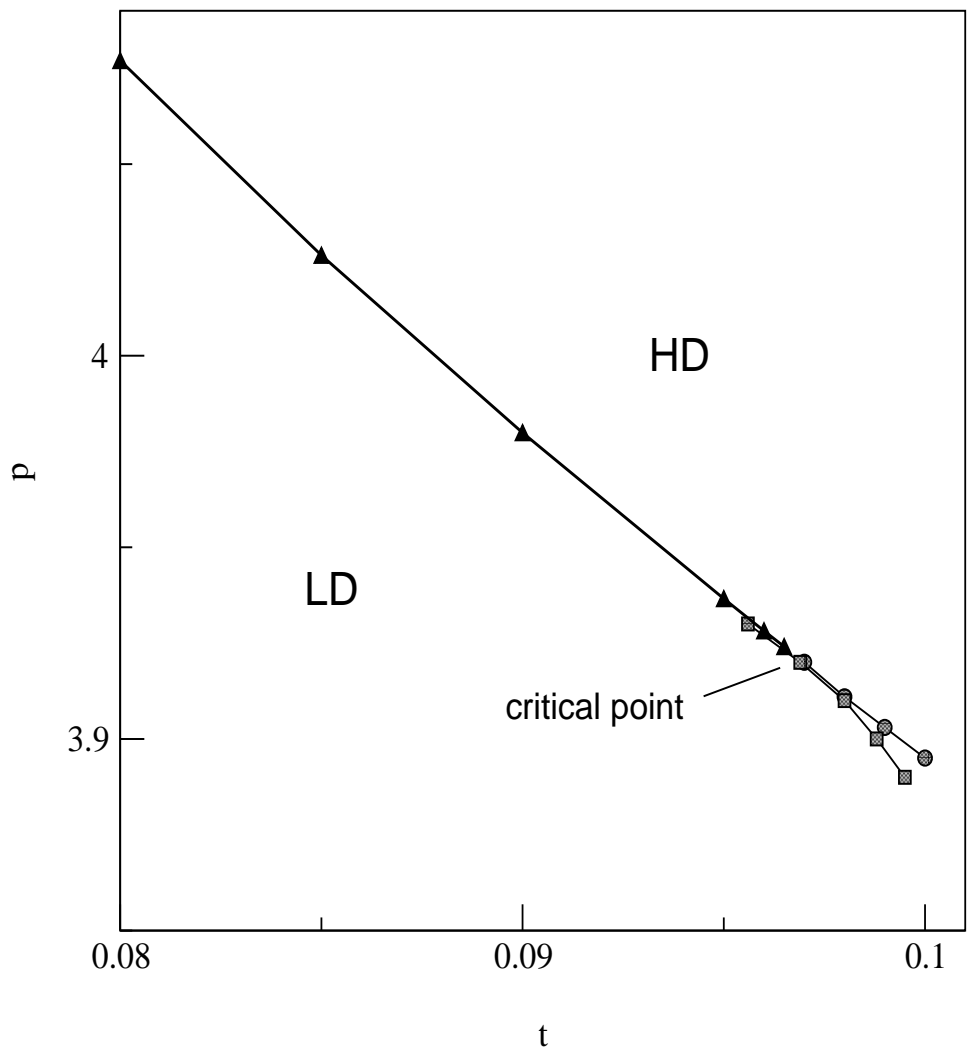}

\vspace*{3cm}

\caption{\footnotesize Lines of specific heat (squares) and
compresibility (circles)
maxima and coexistence line (triangles) near the critical point. \label{compres-calor}}


\end{figure}

Compressibility and  specific heat for the model can be obtained 
numerically. These quantities present maxima whose magnitude increase with
decreasing temperature, as expected on  approach of the critical point. The
line of specific heat and compressibility maxima is shown in Fig.
\ref{compres-calor}. The distance between the two lines grows as pressure is
lowered. The specific heat maxima are localized at lower temperatures with
respect to the compressibility ones, a feature also found for
the model discussed by Sastry and collaborators, which presents no liquid-liquid transition
\cite{sastry}.
  
\section{Liquid-liquid transitions} 

\begin{figure}[th]

\epsfysize=12.77cm \epsfxsize=12.64cm

\epsffile{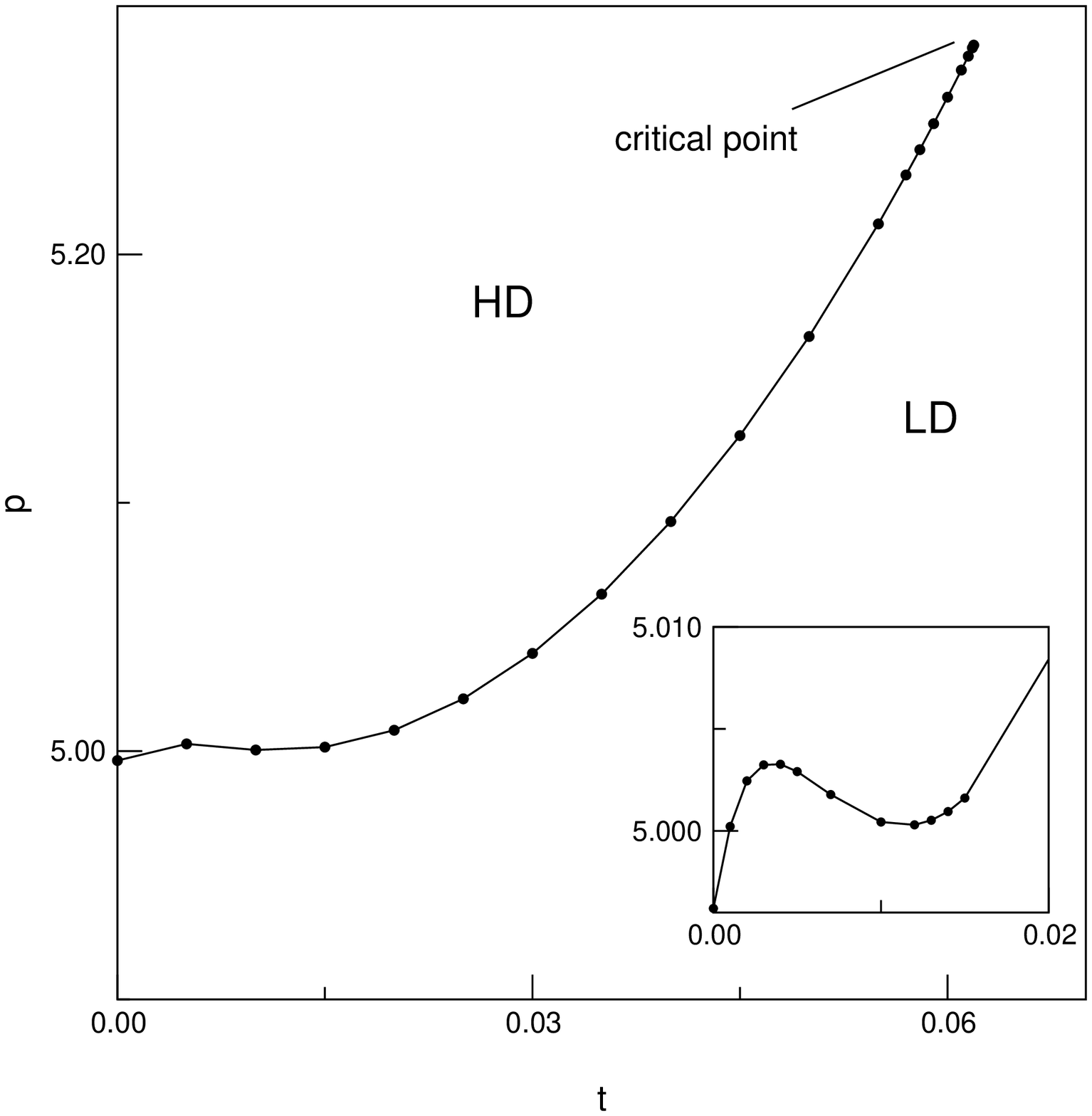}

\vspace*{3cm}

\caption{\footnotesize Coexistence line on the pressure-temperature phase
diagram in the absence of HB net entropy. The
high density phase (HD) has $n_6 > n_4$ and the low density phase (LD)
presents $n_4 > n_6$ for low temperatures and $n_6 > n_4$ for $t>0.06$. 
The critical point is given by $t_c=0.0619$ and $p_c=5.2843$. The inset shows reentrant
behaviour at very low temperatures. \label{coex-sem-ph}}


\end{figure}

In case one does not consider the HB net entropy (Eq. \ref{s-ph}), a different phase
diagram is obtained. As seen in Fig. \ref{coex-sem-ph}, the intermediate density phase
disappears and an inversion occurs with respect to entropy: the higher entropy
phase, in this case, is the low density one, as can be guessed from the slope
of the coexistence curve. These features might be related to other
liquid-liquid transition such as that of carbon.

The different structures of solid carbon, diamond and graphite, are a result
of hybridization of the carbon atoms. 
The diamond structure, stable at higher pressures, is associated to $sp^3$
tetrahedral hybridization, while graphite is constituted of planes of $sp^2$
hybrids. Graphite is less dense, but of higher entropy.
Experimental and theoretical results \cite{exp-c,glosli,thiel} point to
the possibility of a liquid-liquid phase transition, between a high
density liquid phase (dominated by $sp^3$ hybridized atoms) and a low density
liquid phase (dominated by $sp$ hybridized atoms in the case of MD simulations
\cite{glosli} and by $sp^2$ hybridized atoms in a scaling formalism model
\cite{thiel}). As in the solid state \cite{c-phase-diagram}, the slope of
the liquid-liquid coexistence line in the pressure versus temperature diagram is positive
\cite{glosli,thiel} and, according to Le Chatelier's principle, the
lower density phase must present the largest entropy.
In the absence of the HB net entropy, the model we propose could be
thought of as a mean-field treatment of Brenner's potential, used to describe
carbon, and for which bond energies depend on the local environment in such
a way as to produce the correct geometries and energies of the known carbon
structures \cite{glosli,brenner}. Is is not clear whether it would be necessary
to consider additional entropy terms, because of the mean-field treatment.
As a curiosity, in our model, from $t_c=0.062$
(Fig. \ref{coex-sem-ph}), if we consider carbon bond energy
of the order of $600 kJ/mol$ \cite{chang}, we obtain $T_c \simeq 5000 K$ (against
$9000 K$ from MD simulations \cite{glosli}).

The two cases studied lead to a simple description of potential candidates for
liquid polymorphism. We discuss this in terms of ananalysis of minimization
of the Gibbs free-energy. At small
temperatures, at which energy wins out against entropy, pressure {\it may}
produce a transition betweeen a high and a low density {\it if}
the low energy phase is the less dense ($\Delta G \simeq _{small T}
\Delta U - P | \Delta V| < 0$
at some pressure, where $\Delta X$ is defined as $X_{HD}-X_{LD}$).

Thus liquid-liquid transitions may arise if low lying energy states are low
density states (as is the case of water or carbon) and for any model with
energy monotonically increasing with density. Of course, in order for the
system to present also a gaseous phase, mean field or van der Waals energy
should increase towards very low densities, and thus present an
extremum, as pointed out by Tejero and Baus \cite{tejero} and also implied
in Poole and collaborators' proposal \cite{poole}.
However, we understand the van der Waals attractive potential would have to
present a minimum (as can be inferred from \cite{poole}): as density increases
from the gas region, the potential decreases, as in usual van der Waals
theory, but a minimum is present at some intermediate density, above which
the potential slightly increases, destabilizing a very dense liquid phase, as
in this model. Once the condition for the liquid-liquid transition is
fulfilled, two possibilities arise for the slope of the coexistence line, and
these, according to Le Chatelier's principle, will depend entirely on
entropy: if the high density phase is the higher entropy phase, as in water,
we encounter a negative slope, whereas if it is a smaller entropy phase, as
in carbon, we encounter a positive slope. 
 
\section{Discussion and summary}    
 
In liquid water there are two sources of disorder, not really
independent: positional disorder of the molecules and disorder of hydrogen
bonds. In the case of the latter, molecules may loose HBs either (i)
because there are no nearby molecules free to engage in an HB (as in the
case of a disfavourable increased local coordination - see Fig. \ref{desenho}-b) or (ii)
because of rotations which remove the necessary alignment between
neighbouring molecules.  In some of the models proposed in the literature
only the second feature has been considered \cite{nadler,square}, in terms of
ice-like models, while in others \cite{sastry,sastry-model,debe} molecular
rotations leading to weak bonds are described through Potts variables on cubic
lattices, and smaller volume attributed to the disordered bonds in order to
account for the first feature (increased non-alignment in dense
environments). Interestingly, both kinds of model (except \cite{debe})
present only a liquid-gas transition and no low-temperature transition, in
support of the ``singularity-free" scenario. A low-temperature transition in
the ice model would imply a discontinuity in the number of HBs at the
transition, a requirement which would also have to be met in the Potts models
(except \cite{debe}), for which volume is measured directly in terms of
bonds. 

In the case of the model discussed in this paper, only the first feature, the
distortion of the HB net, leading to rupture of HBs, not due to non-alignment,
but to disponibility of receptor or donor molecules, is taken into account. In
this case, liquid-liquid transitions are present, independently of model
parameters, but for specified geometric constraints for local structure.

In relation to the dispute on the existence of two supercooled waters , the
answer could depend on which of the two mechanisms above, restricted bonding
due to varying local coordination, or weak bonding due to thermally induced
non-alignment, is most relevant.

Our study of the Bernal model for liquids on the pressure-temperature
plane has shown the possibility of liquid-liquid coexistence if energy
favours low density configurations. Liquid polymorphism both above and
below the fusion line have been suggested in the literature for
associating, as well as for non-associating liquids. Our results point
to the role of bond network entropy. The latter becomes  patent if one
compares Figs. \ref{coex} and \ref{coex-sem-ph}. The inclination of the
coexistence lines present opposite signs in the critical point region,
for the two systems.

The compressibility and specific heat maxima beyond the critical
temperature could relate to the so called anomalous properties of
water. However, as for other models discussed in the literature, it
is senseless to discuss whether the polymorphisms belongs to
supercooled or stable liquid phase, since it is unable to describe the
solid phase. Also, the model does not allow for ``holes'' and is
therefore uncapable of describing the liquid-gas transition, but this
is not really a shortcoming, because interest lies really in the dense
phases.
                      
\section{Acknowledgments}

We thank S. R. Salinas and M. J. Oliveira for indicating references on
supercooled water. One of us (NG) aknowleges financial support from
Fapesp.

\newpage

\newpage

\newpage

\newpage

\newpage

\begin{references}
\bibitem{foot}This choice is dictated by the fact that coordination
four defines the loosest packed lattice and not by the real water property, relevant in three dimensions.
\end{references}
\end{document}